\documentclass[twocolumn]{aa} 
\usepackage{hyperref}
\usepackage{dblfloatfix}
\usepackage{txfonts}
\usepackage{CJK}
\usepackage{booktabs}
\usepackage{graphbox}
\usepackage{amsmath}
\usepackage{soul}
\usepackage{scrextend}
\deffootnote[0.45em]{0.2em}{0.2em}{\textsuperscript{\thefootnotemark}\,}
\usepackage{amsmath,url}
\usepackage{algorithm}
\usepackage{amsfonts}
\usepackage{mathrsfs}
\usepackage{bm}
\usepackage{rotating}
\usepackage{color}
\usepackage{graphicx}
\usepackage{subfigure}
\usepackage[toc,page]{appendix}
\usepackage[english]{babel}
\usepackage{orcidlink}
\usepackage[mathlines,switch]{lineno}
\newcommand\gaia{\textit{Gaia~}}
\newcommand\gdr[1]{\gaia~DR#1}

\newcommand\masyr{\ensuremath{\text{mas~yr}^{-1}}}

\newcommand\kms{\ensuremath{\text{km~s}^{-1}}}

\newcommand\gband{\ensuremath{G}}
\newcommand\gbp{\ensuremath{G_\mathrm{BP}}}
\newcommand\grp{\ensuremath{G_\mathrm{RP}}}
\newcommand\RAdeg{\ensuremath{\text{$\alpha$}}}
\newcommand\DEdeg{\ensuremath{\text{$\delta$}}}
\newcommand\plx{\ensuremath{\text{$\varpi$}}}
\newcommand\pmra{\ensuremath{\text{$\mu_{\alpha}^*$}}}
\newcommand\pmdec{\ensuremath{\text{$\mu_{\delta}$}}}

\newcommand\kmskpc{\ensuremath{\kms~\text{kpc}^{-1}}}
\newcommand\Gconst{\ensuremath{\text 4.3 \times 10^{-6}~\text{kpc}~({\kms})^{2}~{M_{\odot}^{-1}}}}

\begin{document} 
\begin{CJK*}{UTF8}{gbsn}
   \title{Formation and evolution of new primordial open cluster groups: Feedback-driven star formation}
   \author{Guimei Liu (刘桂梅)\inst{1,2}\orcidlink{0009-0002-4686-8115}
          \and Yu Zhang (张余)\inst{1,2}\orcidlink{0000-0001-7134-2874}
          \and Jing Zhong (钟靖)\inst{3}\orcidlink{0000-0001-5245-0335}
          \and Li Chen (陈力)\inst{3,2}\orcidlink{0000-0002-4907-9720}\\
          \and Xiangcun Meng (孟祥存)\inst{4,5}\orcidlink{0000-0001-5316-2298}
          \and Kai Wu (吴开)\inst{6}\orcidlink{0000-0003-0349-0079}
          }
   \institute{XinJiang Astronomical Observatory, Chinese Academy of Sciences,150 Science 1-Street, Urumqi, Xinjiang 830011, China  \\ \email{zhy@xao.ac.cn}
   \and School of Astronomy and Space Science, University of Chinese Academy of Sciences, No. 19A, Yuquan Road, Beijing 100049, China
            \and Astrophysics Division, Shanghai Astronomical Observatory, Chinese Academy of Sciences,80 Nandan Road, Shanghai 200030, China\\ \email{jzhong@shao.ac.cn}
            \and Yunnan Observatories, CAS, P.O. Box 110, Kunming 650216, Yunnan, China
            \and International Centre of Supernovae, Yunnan Key Laboratory, Kunming 650216, China
            \and Astronomisches Rechen-Institut, Zentrum f\"{u}r Astronomie der Universit\"{a}t Heidelberg, M\"onchhofstr. 12-14, D-69120 Heidelberg, Germany}
   \date{Received ..., 2024; accepted ...}

  \abstract
   {Open cluster (OC) groups are collections of spatially close OCs originating from the same giant molecular cloud. The formation mechanism of the OC groups remains unclear due to limited sample size and data precision. Recent advances in astrometric data from the \gaia\ mission have provided an unprecedented opportunity to study OC groups in detail.} 
   {We aim to extend the sample of OC groups and explore the formation and evolution mechanisms of the newly identified primordial OC groups. We focus on the impact of stellar feedback events occurring near the OC groups and their role in triggering star formation within these groups.}
   {We identified the OC group using \gaia data by analyzing the close correlations in three-dimensional (3D) spatial, 3D velocity, and age. We conducted $N$-body simulations to trace the dynamical evolution of these groups and obtained the birthplace of OCs. A region where supernova (SN) explosions most likely occurred was predicted around the birthplaces of OC groups, based on the correlation between OC ages and their separation from the possible SN explosion sites. We also traced the orbits of pulsars (PSRs) using the Galactic potential model to probe their association with predicted SN explosion regions.}
   {We report the detection of four OC groups. The member OCs within each group are spatially proximate and exhibit similar velocities. The age spread of these OC groups is within 30 Myr, consistent with the duration of continuous star formation events. Dynamical simulations show that these OC groups gradually disperse over time, eventually evolving into independent OCs. The inference can be made that there exist specific regions surrounding Group 1 and Group 2 where the occurrence of  SN explosions is highly probable.
   The strong correlations between OC ages and their separation from predicted SN explosion sites reveal a notable age gradient outward from the SN explosions.
   Additionally, we detected three PSRs near Group 1 and 26 PSRs near Group 2, whose birthplaces align with the predicted SN explosions regions.}
  {The member OCs within each OC group originate from the same molecular cloud, forming through a process of sequential star formation.
  We propose a star formation scenario in which multiple SN explosions triggered the formation of Group 1 and Group 2. Our results support the supernova-triggered star formation process and also reinforce the hierarchical star formation model, highlighting the multi-scale interactions that drive star and open cluster formation.}

   \titlerunning{Feedback-driven star formation}
   \authorrunning{Guimei, Liu, et al.}
   \maketitle

\section{Introduction}
Open clusters (OCs) are stellar systems consisting of dozens to thousands of stars that form from the same molecular cloud and remain gravitationally bound \citep{Lada2003}. While most OCs exist as independent stellar systems, some form in groups, such as binary clusters \citep{2019A&A...624A..34Z,Qin2025}  or more complex configurations involving multiple OCs \citep{Conrad2017, Bica2003A&A405991B, Camargo2016, Casado2021}. These OC groupings are spatially close and share a common origin from the same giant molecular cloud (GMC), known as cluster complexes \citep{Piskunov2006} or primordial groups \citep{Conrad2017}. In this paper, we use the term "OC groups."
OC groups are large-scale structures that extend over several hundred parsec (pc). They consist of many substructures on smaller scales, ranging from a few to tens of pc \citep{Efremov+1978}. Star formation in OC groups takes place over a span of several Myr to tens of Myr \citep{Casado2021, Casado2022Univ8113C}.

There are two major scenarios for the formation of OCs. The monolithic model \citep{Lada1984ApJ} suggests that clusters form in the densest parts of molecular clouds, where a single, intense star-formation event creates a compact cluster within its primordial gas. Feedback processes later expel gas, causing the star cluster to expand and stars to eventually disperse into the field \citep{Lada2003}. The hierarchical model \citep{Kruijssen2012} proposes that high-density regions with high star formation efficiency form bound star clusters, while lower-density regions remain substructured and unbound after gas expulsion. Determining which theories govern the formation of OCs remains a matter of ongoing debate.
The formation of OC groups provides strong support for the hierarchical cluster formation theory \citep{Cantat2019ring, Grossschedl2021, Wang2022, Ratzenb+2023}, as OC groups likely form through sequential processes within GMCs. Thus, studying OC groups offers deeper insights into hierarchical star formation modes within GMCs \citep{Bonnell2003}. Analyzing the interactions between member OCs within an OC group reveals their internal dynamical evolution, enhancing our understanding of the mechanisms governing the formation and evolution of OCs.

Before the \gaia mission, constructing a reliable census of OCs was challenging due to the limitations in astrometric data, which significantly hindered the identification of OC groups. The absence of a complete OCs sample and precise kinematic data made it difficult to accurately determine the spatial and kinematic relationships between OCs.
Since 2018, \textit{Gaia} has provided vast amounts of high-precision astrometric parameters (\RAdeg, \DEdeg, \plx, \pmra, \pmdec) and multi-band photometric data like \gband, \gbp, \grp\ \citep{gaiadr2,gaiaEdr3, GaiaDR3_2022}, along with an unprecedentedly large sample of stellar radial velocity (RV) \citep{Katz2019, Katz2023}. Coupled with the widespread adoption of machine learning algorithms has led to the discovery of numerous new OCs within the Milky Way \citep{Cantat-Gaudin2019b624, CastroGinard_2020, Liu_2019, 2021RAA....21...45Q, Qin_2023ApJS},  while known OCs have also been found to have more member stars and larger spatial scales \citep{ 2019A&A...624A..34Z, 2020ApJ...889...99Z,2022RAA....22e5022B, 2022AJ....164...54Z}.
Based on these advances, identifying primordial OC groups has become a growing area of research. 
Numerous studies have been conducted on OC groups over several decades. 
These newly discovered OC groups exhibit diverse spatial distributions and have been given intriguing names such as "rings" \citep{Cantat2019ring}, "strings" \citep{Kounkel2019_3kpc}, "relic filaments" \citep{Beccari2020}, "snakes" \citep{Tian2020}, "Huluwa" \citep{Pang2021}, and "pearls" \citep{Coronado2022}. 
These results offer new insights into how OC groups form and evolve and provide observational evidence supporting the hierarchical star formation model.

From a theoretical perspective, star formation is a complex process influenced by numerous factors, including feedback mechanisms such as stellar wind, radiation pressure, and SN explosion \citep{Swiggum2021}. The feedback from massive stars regulates the star formation efficiency and shapes the environment in which the stars form. 
Hierarchical star formation induced by stellar feedback operates through two primary mechanisms. Firstly, feedback from the first generation of massive stars (stellar wind, radiation pressure) drives molecular cloud collapse and subsequent star formation \citep{Elmegreen+2000, Kerr+2021, MacLow2004RvMP, Fujii+2021b}. This externally perturbation-driven star formation typically occurs along filamentary molecular cloud structures.
The second mechanism involves shockwaves from supernovae (SNe) compressing the surrounding gas and creating high-density regions that trigger new star formation. This type of star formation typically occurs at the edges of shells formed by SNe \citep{Kounkel2020Apj, Ehlerov+2018}. Among these feedback mechanisms, SNe are the primary source of energy injection into the local interstellar medium (ISM). In contrast, stellar winds, being a continuous process, exert a smaller but cumulative pressure that also promotes star formation \citep{MacLow2004RvMP}.

In recent years, multiple numerical simulations have investigated how the supernova (SN) event influences star formation. Overall, these investigations have revealed a common pattern: whether to be an individual SN event \citep{Ostriker+2015, Li2015ApJ, Martizzi2015MNRAS} or a sequential outburst from multiple SNe \citep{Padoan+2016}, the majority of the imparted kinetic energy is transferred to the low-density regions of the interstellar medium (ISM). Observational data provides compelling evidence that SNe indeed trigger the collapse of molecular clouds in their surrounding environment, leading to star formation \citep{Ehlerov+2018, Kounkel2020Apj, Zucker2022Nature, Swiggum2024Natur}, with an impact scale reaching from hundreds of pc to kiloparsecs (kpc; \citealt{Stahler2004book}).  
Despite the growing theoretical and observational evidence, the mechanisms underlying OC formation remain unclear, particularly the influence of feedback from massive stars.

Our previous work \citep{Liu2025} identified four new primordial OC groups. Here, we propose a possible formation mechanism for two specific OC groups. This paper is structured as follows:
In Sect.~\ref{sec:paper1}, we summarize the data and methods from our previous work, including the data sources for the OC sample, the methods used to determine OC parameters, and the search methodologies.
In Sect.~\ref{sec:Discussion}, we analyze the properties of the OC groups and discuss the evolutionary outcomes of the four OC groups, as well as the formation mechanisms of G1 and G2. Finally, we summarize our findings in Sect.~\ref{sec:summary}.

\begin{table*}[ht]
\centering 
\caption{Property parameters of member OCs.} 
\label{tab:OC_xzy} 
\footnotesize 
\setlength{\tabcolsep}{5.0pt} 
\renewcommand{\arraystretch}{1.5} 
\begin{tabular}{rrrrrrrrr}
\hline \hline 
 Group& Cluster         & X        & Y       & Z       &U    & V     & W    & Mass   \\
 ~ & ~ & \ensuremath{\text{(pc)}} & \ensuremath{\text{(pc)}} & \ensuremath{\text{(pc)}} & (\kms) & (\kms) & (\kms) & \ensuremath{\text{(M$_{\odot}$)}} \\ 
     (1) & (2) & (3) & (4) & (5) & (6) & (7) & (8) & (9) \\ 
    \hline \hline
G1 & ASCC 105     & -8458.28 & -79.93  & -293.93 & 26.81 & 234.50  & 12.47 & 120.49 \\
 & OCSN 113     & -8308.90  & -86.88  & -174.61 & 28.39 & 233.57 & 14.12 & 263.00  \\
 & OCSN 117     & -8278.64 & -86.82  & -138.26 & 27.83 & 232.37 & 15.23 & 147.45 \\
 & OCSN 152     & -8377.73 & -56.72  & -194.30  & 29.78 & 233.07 & 13.72 & 86.46  \\
 & Roslund 5    & -8461.93 & -142.88 & -236.78 & 28.17 & 234.91 & 12.72 & 164.87 \\
 & UPK 82       & -8433.32 & -98.38  & -293.34 & 27.46 & 233.28 & 11.53 & 83.93  \\ \hline
G2 & OCSN 16      & -8332.08 & 4.90     & -153.54 & 27.67 & 234.58 & 8.64  & 124.37 \\
 & OCSN 18      & -8337.29 & 8.35    & -139.57 & 28.22 & 235.18 & 9.47  & 62.80   \\
 & Stephenson 1 & -8329.54 & -3.54   & -121.11 & 29.08 & 234.74 & 9.06  & 120.94 \\
 & UPK 101      & -8330.22 & -14.24  & -103.59 & 28.89 & 234.22 & 10.71 & 194.28 \\
 & UPK 64       & -8358.74 & 16.17   & -221.86 & 28.35 & 238.63 & 8.44  & 127.69 \\
 & UPK 72       & -8385.73 & 8.69    & -229.73 & 29.01 & 238.48 & 8.19  & 77.81  \\
 & UPK 78       & -8334.40  & 26.29   & -147.70  & 27.95 & 235.63 & 8.69  & 55.70   \\
 & UPK 83       & -8316.93 & -6.80    & -139.64 & 27.07 & 233.31 & 8.59  & 42.68  \\  \hline
G3 & OCSN 13      & -8384.72 & -30.85  & -302.08 & 27.73 & 235.44 & 0.75  & 268.57 \\
 & OCSN 14      & -8405.55 & -31.51  & -299.38 & 22.14 & 235.10  & -1.76 & 48.53  \\
 & UPK 70       & -8397.83 & -6.92   & -276.91 & 27.09 & 238.27 & 4.02  & 90.72  \\  \hline
G4& OCSN 22      & -8293.58 & -55.59  & -51.61  & 23.54 & 232.25 & 12.52 & 82.87  \\
 & OCSN 23      & -8320.05 & -106.69 & -60.25  & 17.10  & 232.45 & 11.98 & 142.09 \\
 & RSG 5        & -8309.85 & -97.29  & -75.69  & 17.54 & 231.34 & 11.63 & 167.04 \\ \hline
\end{tabular}
\tablefoot{Column (1): OC group name in this work. Column (2): OC name in this work. Columns (3)–(8): 3D positions and 3D velocities in Galactocentric coordinates based on the right-handed coordinate system. Column (9): Total mass of OC.}
\end{table*}

\begin{figure*}[htb]
    \centering
    \includegraphics[width=0.8\linewidth]{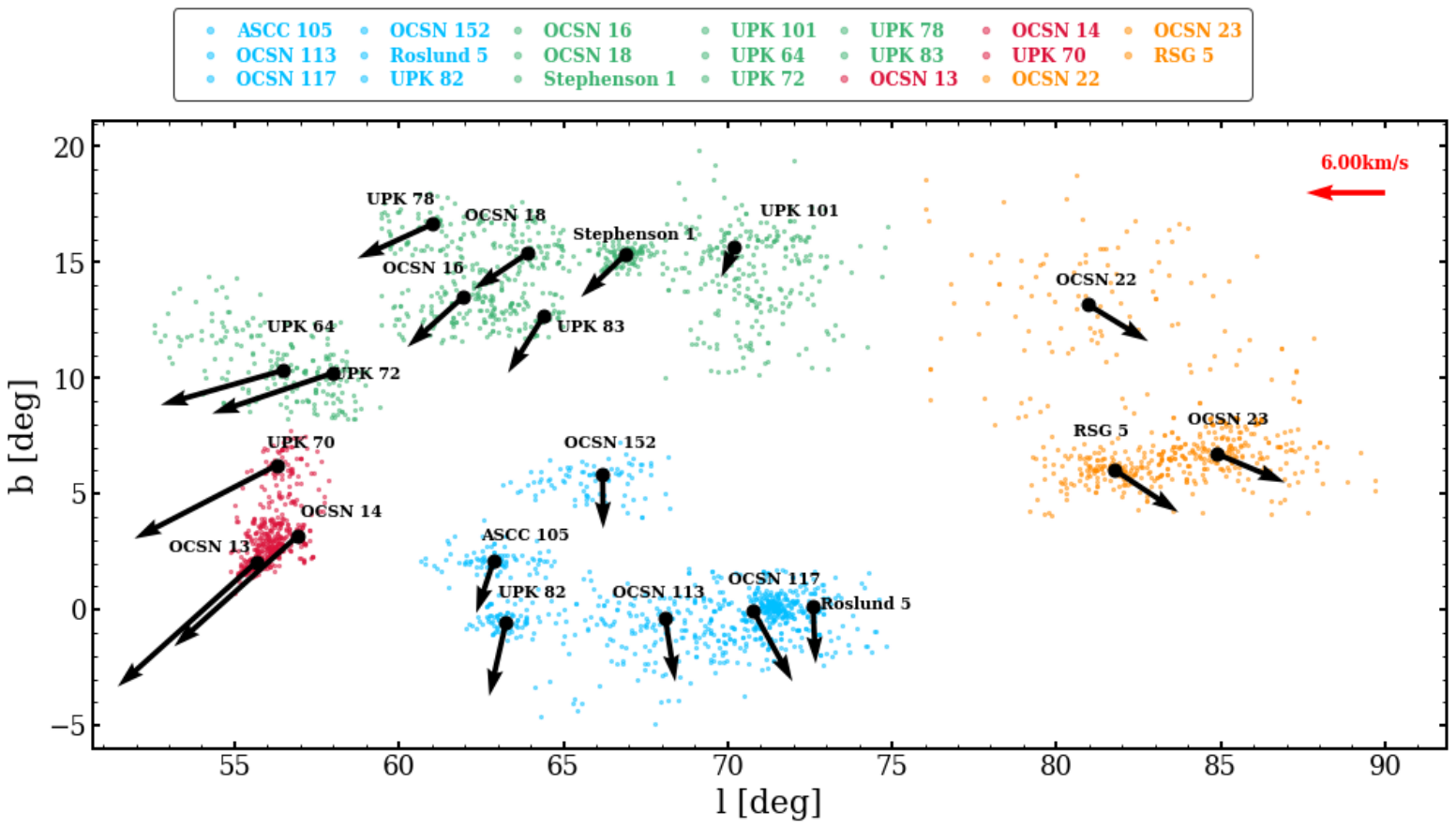}
    \caption{Spatial distribution of OC groups in $l$-$b$. The blue, green, red, and orange dots represent OC groups G1, G2, G3, and G4, respectively. The black arrows denote the tangential velocity of the member OC, with arrow lengths scaled proportionally according to the red reference arrow located in the upper right corner of the figure.}
    \label{fig:lb_Vt}
\end{figure*}

\begin{figure*}
    \centering
    \includegraphics[width=0.8\linewidth]{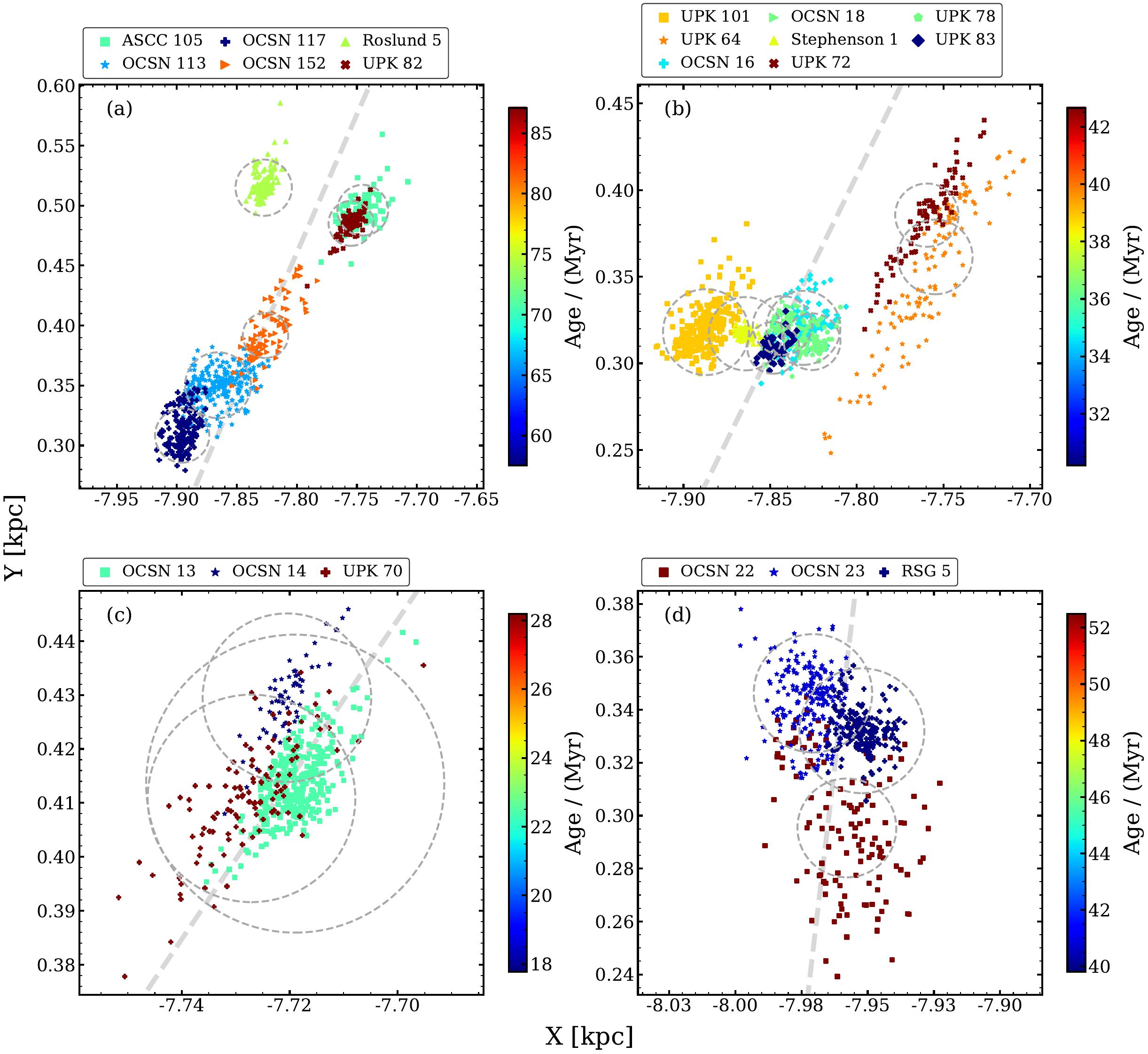}
    \caption{
    Spatial distribution of OC groups after distance correction. Panels (a) G1, (b) G2, (c) G3, and (d) G4 depict the different groups, with symbols representing member stars of the various OCs (color-coded with age as represented in the color bar). The gray dashed lines indicate the lines of sight, while the gray dashed circles represent the locations of 3$r_{\rm t}$ for each cluster.}
    \label{fig:3D_xy}
\end{figure*}

\begin{figure*}[!hbp]
    \centering
    \includegraphics[width=0.8\linewidth]{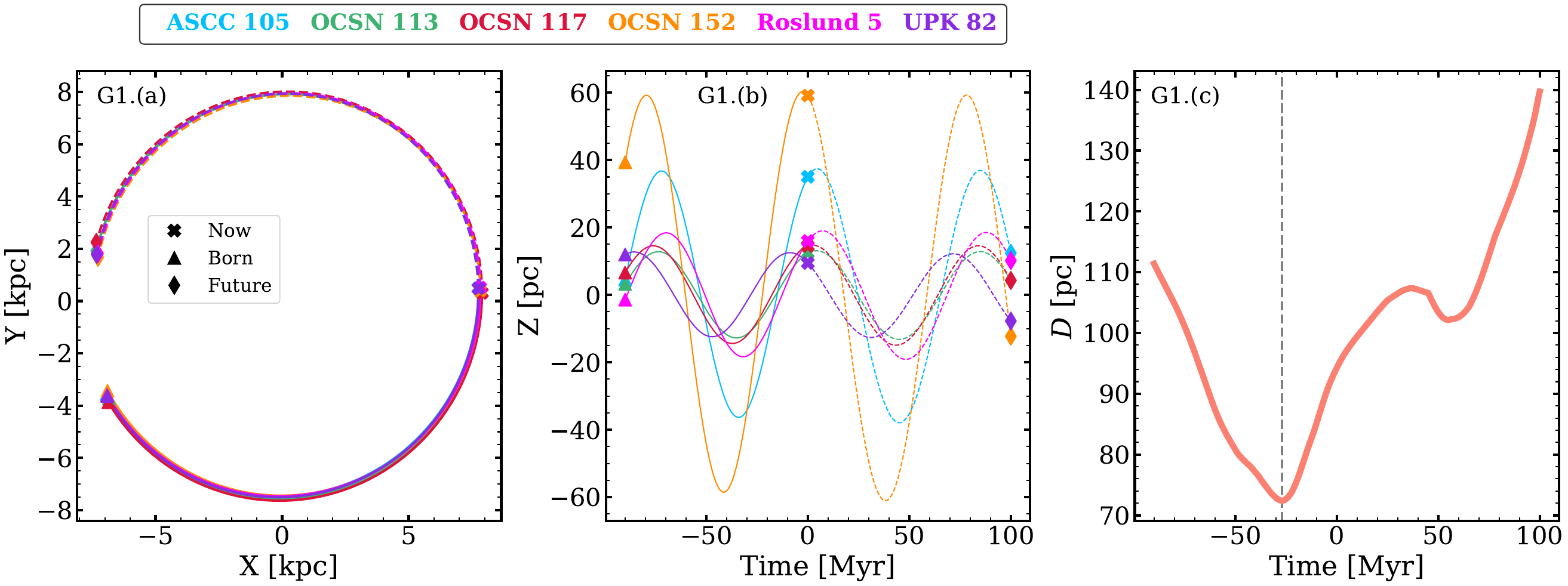}
    \includegraphics[width=0.8\linewidth]{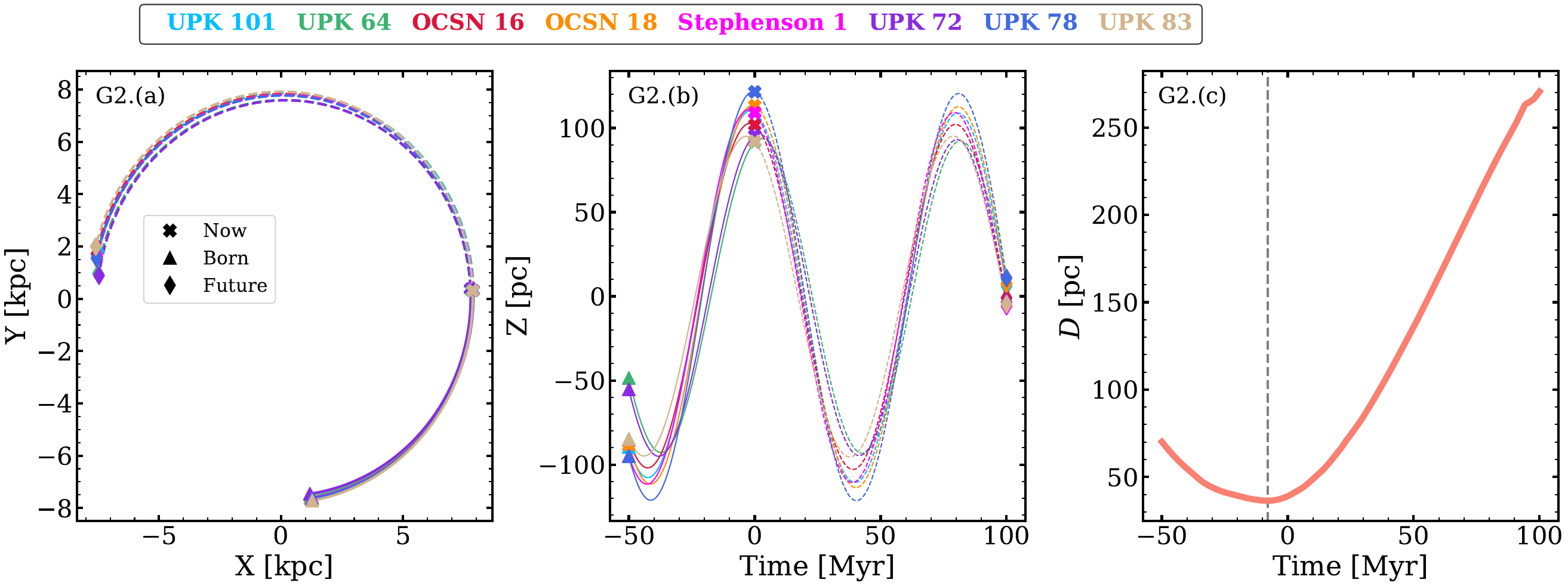}
    \includegraphics[width=0.8\linewidth]{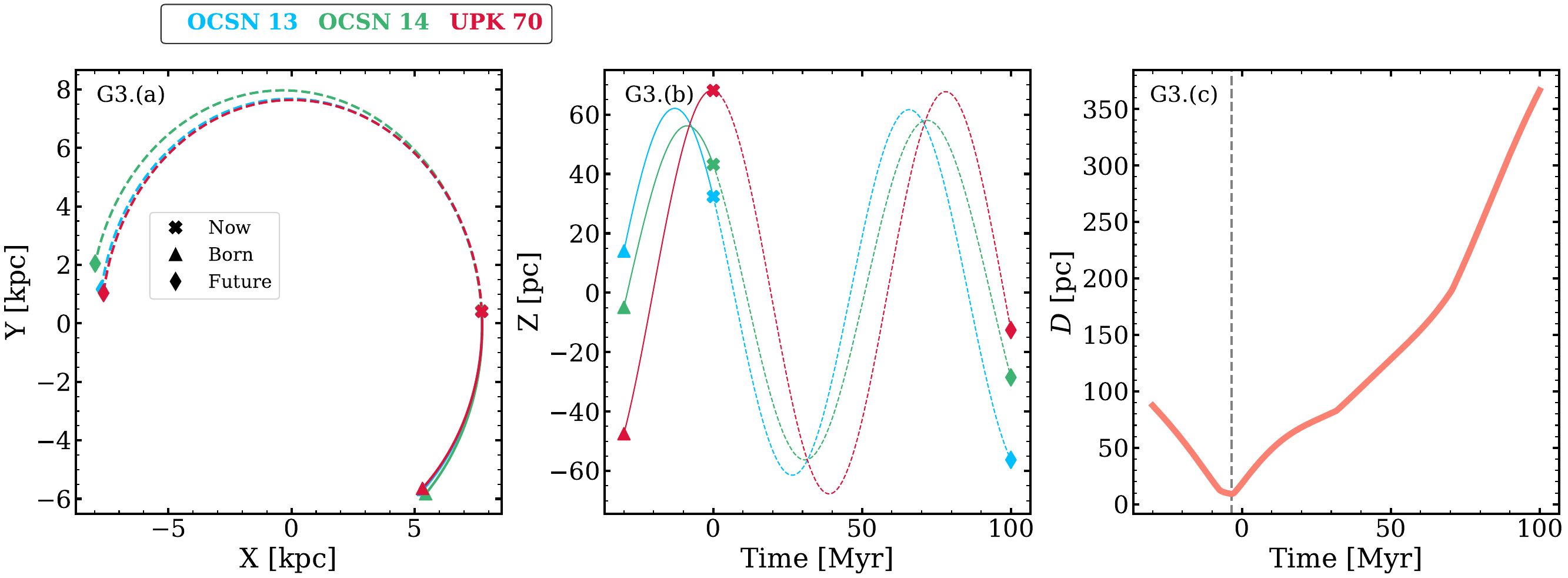}
    \includegraphics[width=0.8\linewidth]{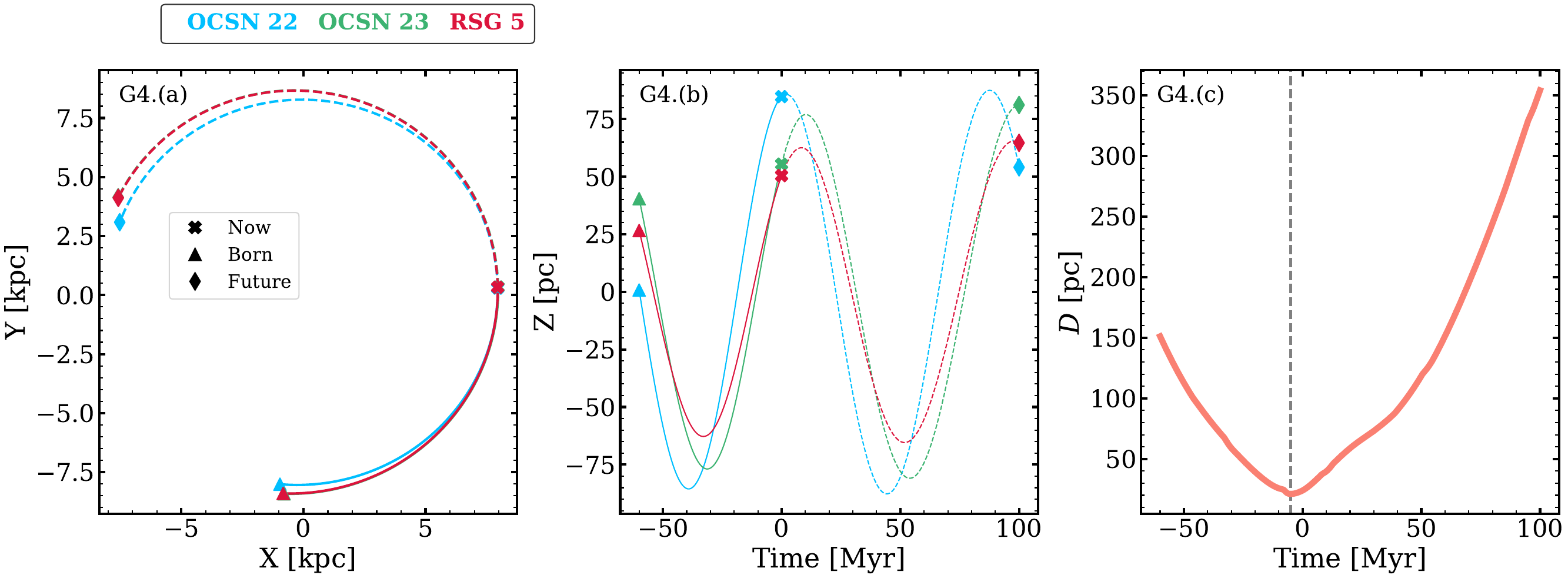}
    \caption{Orbital integration and mean separation between member OCs and the OC group center over time. Panels (a)-(b): Orbital trajectories in the $X$-$Y$ and in $Z$-coordinate over time. Colors are used to distinguish different OCs. Cross symbols mark the present-day locations, triangles represent the past positions of each OC at the birth time of the oldest OC, and diamond symbols denote the predicted positions in 100 Myr.
    Panel (c): The vertical dashed line marks the moment when the separation reaches a minimum. }
    \label{fig:orbit_result}
\end{figure*}
\begin{figure*}[!htb]
    \centering    
    \includegraphics[width=0.75\linewidth]{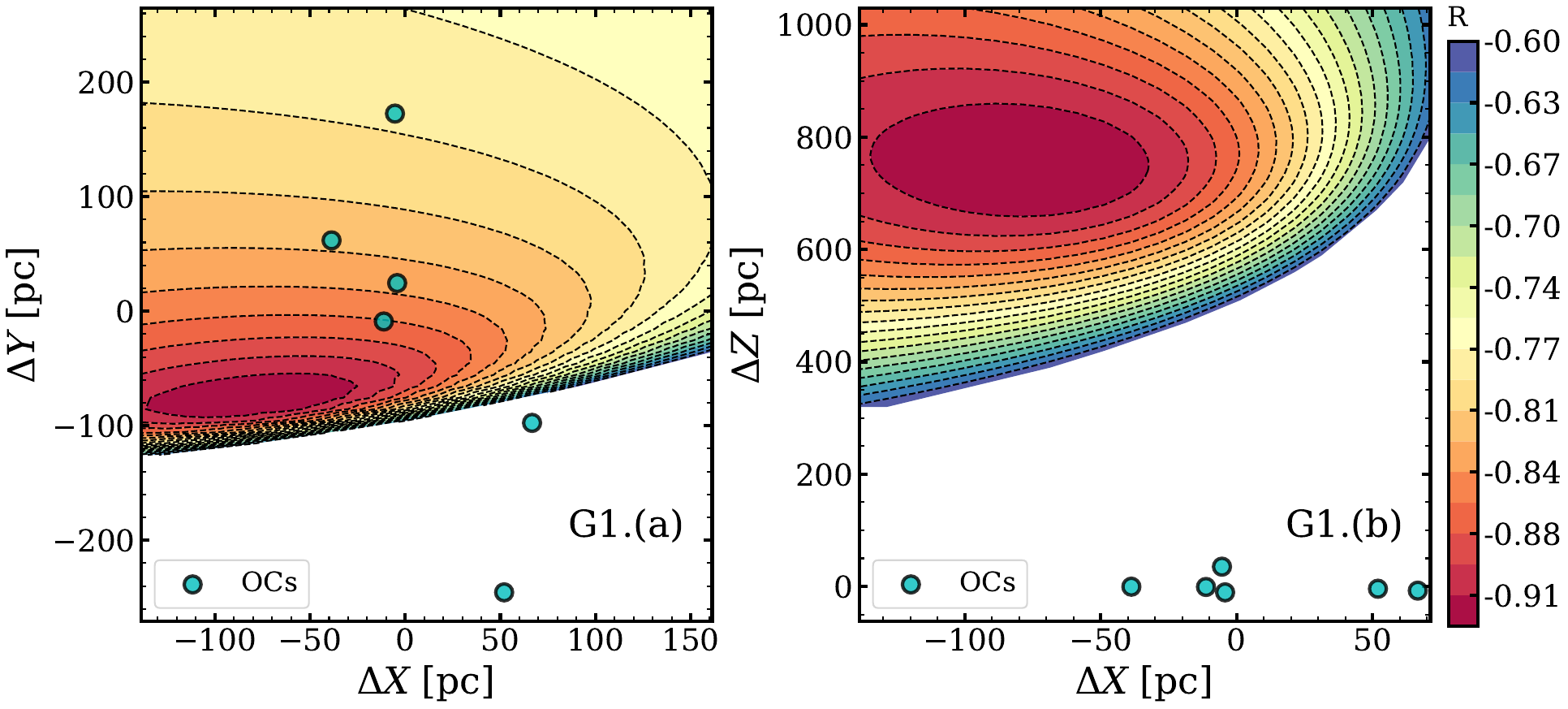}
    \includegraphics[width=0.75\linewidth]{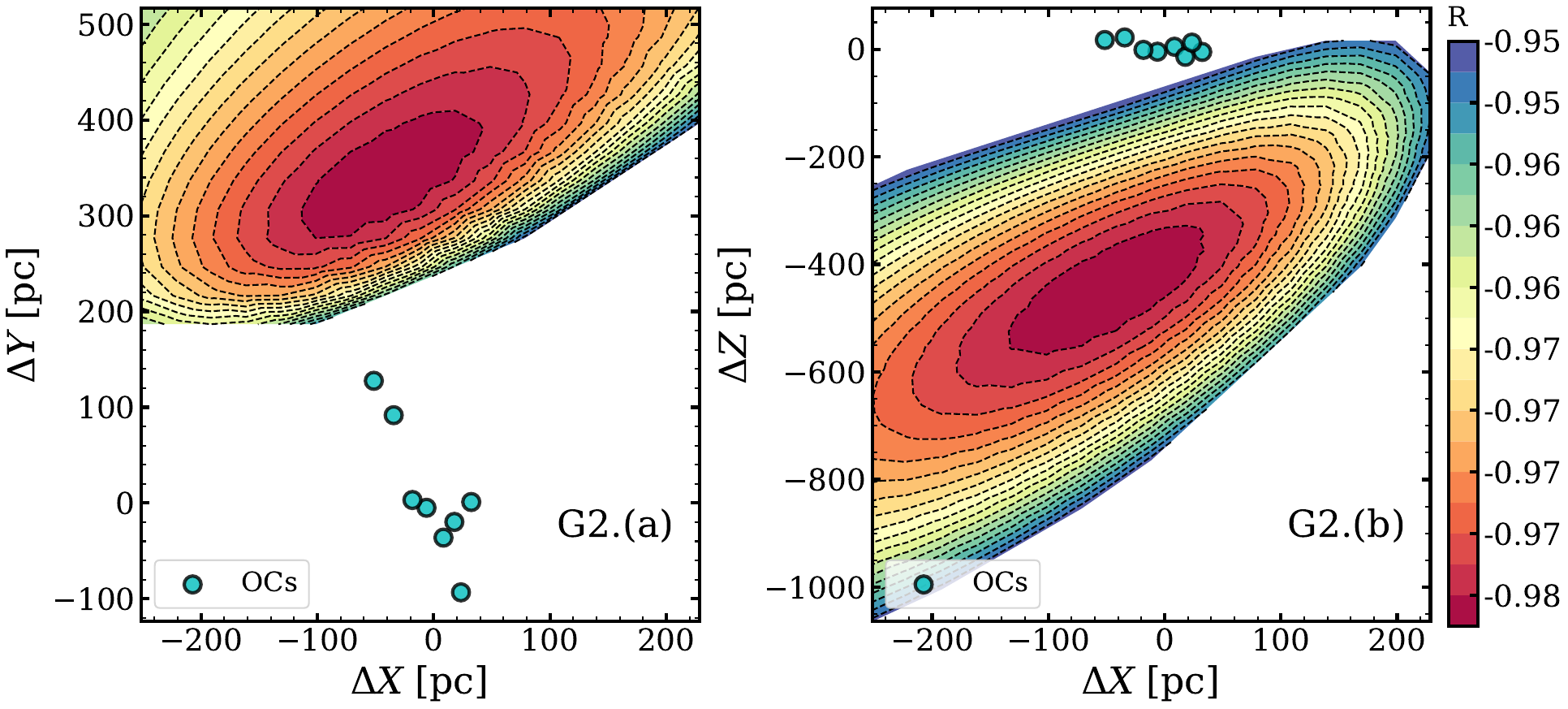}
    \caption{
    Relative 3D spatial distribution of potential SN explosion regions. The cyan dots represent the relative distribution of member OCs at the birth time of the oldest clusters within each OC group. The contour colors represent levels of the correlation coefficient, as shown in the color bar. }
    \label{fig:SNe-mock}
\end{figure*}
\begin{figure*}[!htb]
    \centering
    \includegraphics[width=0.48\linewidth]{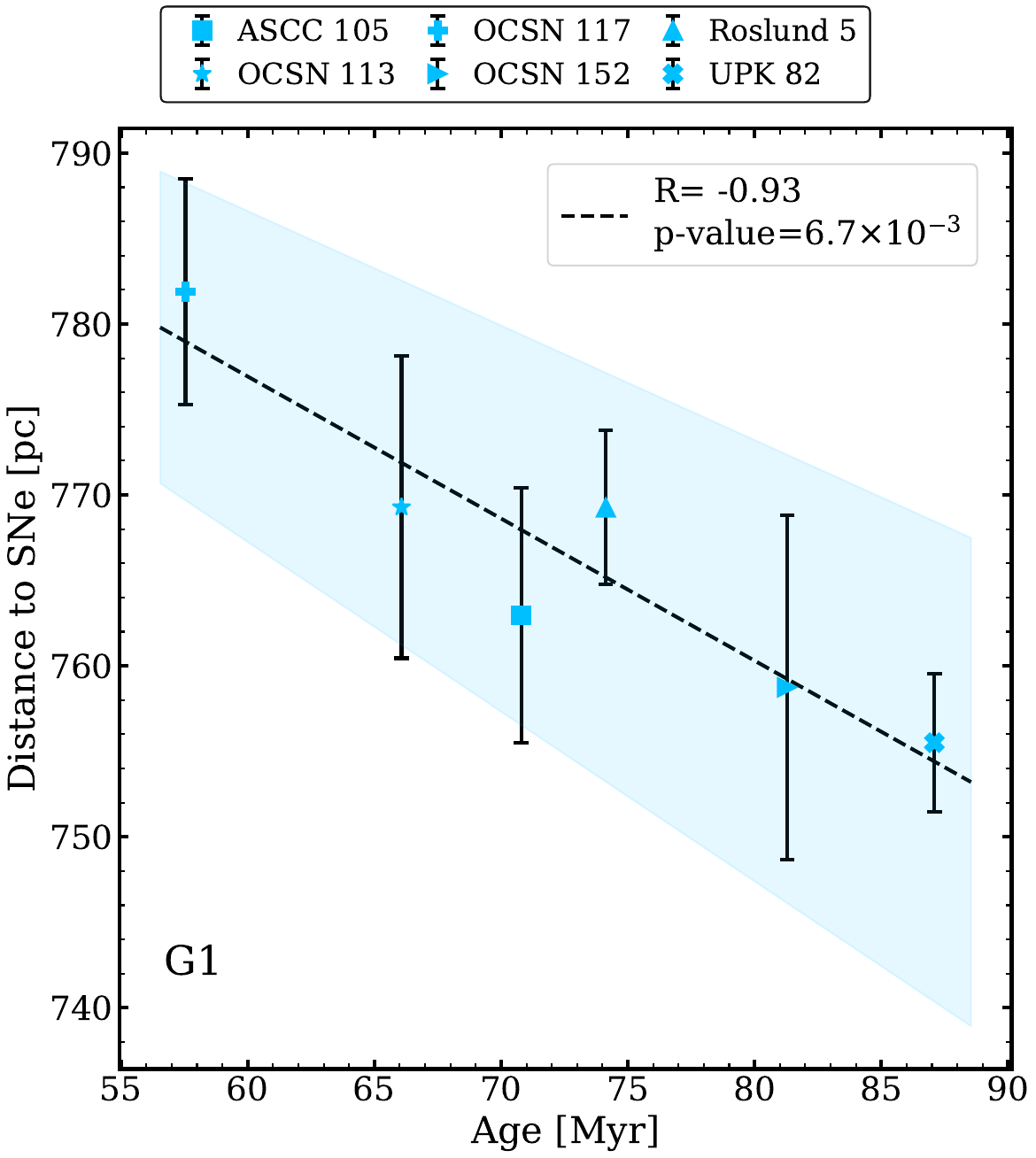}
    \includegraphics[width=0.48\linewidth]{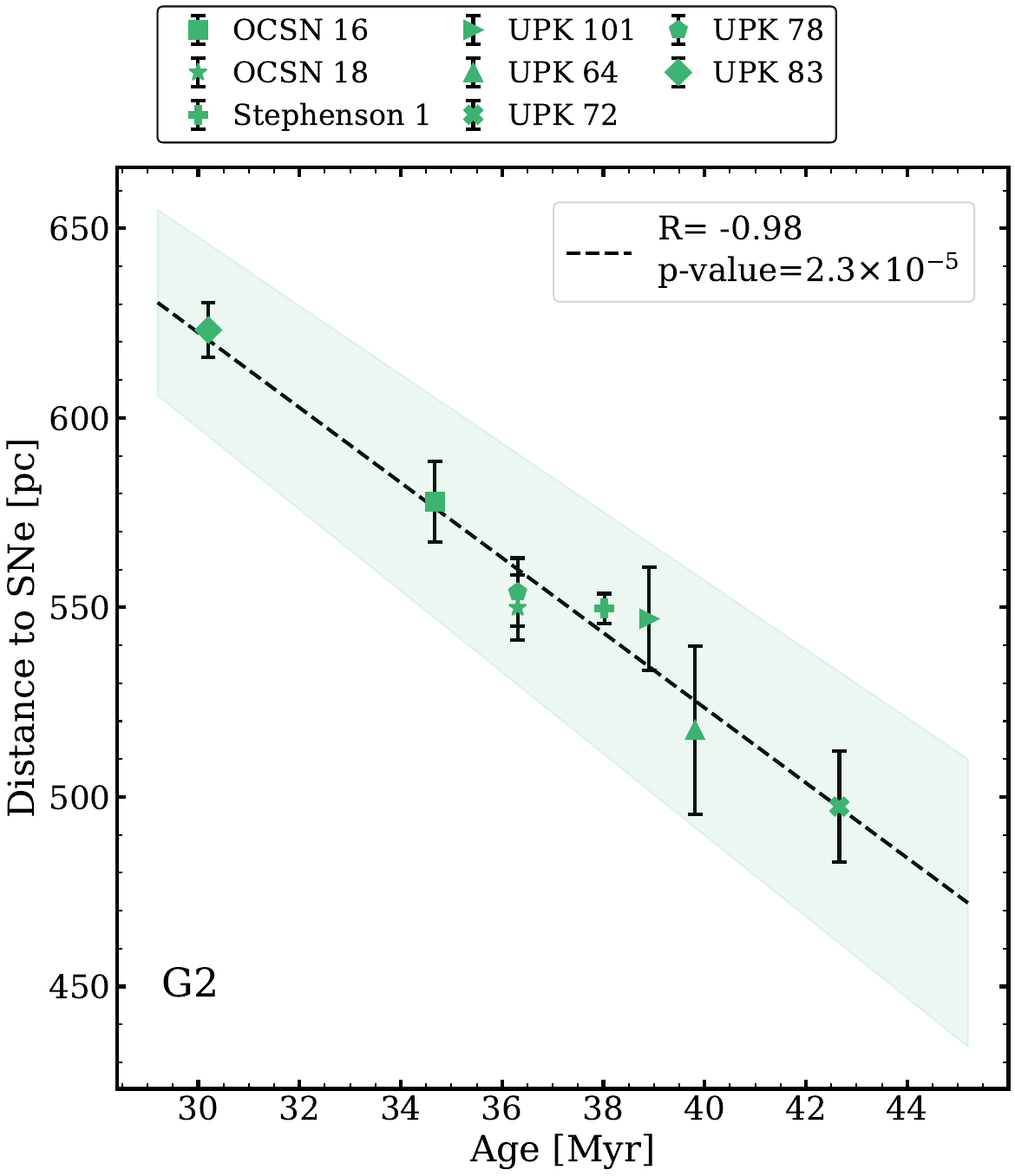}
    \caption{OC ages as a function of the separation to the most probable SN explosion sites. The black dash lines represent the best-fitted linear relation. The left panel shows the result for G1, while the right panel displays the result for G2. The shaded area represents the range of the standard error for the estimated slope. Error bars are computed using error propagation, based on the standard error of the spatial distribution of the member stars.}
    \label{fig:age-Sep}
\end{figure*}

\begin{figure}[!htb]
    \centering
    \includegraphics[width=0.86\linewidth]{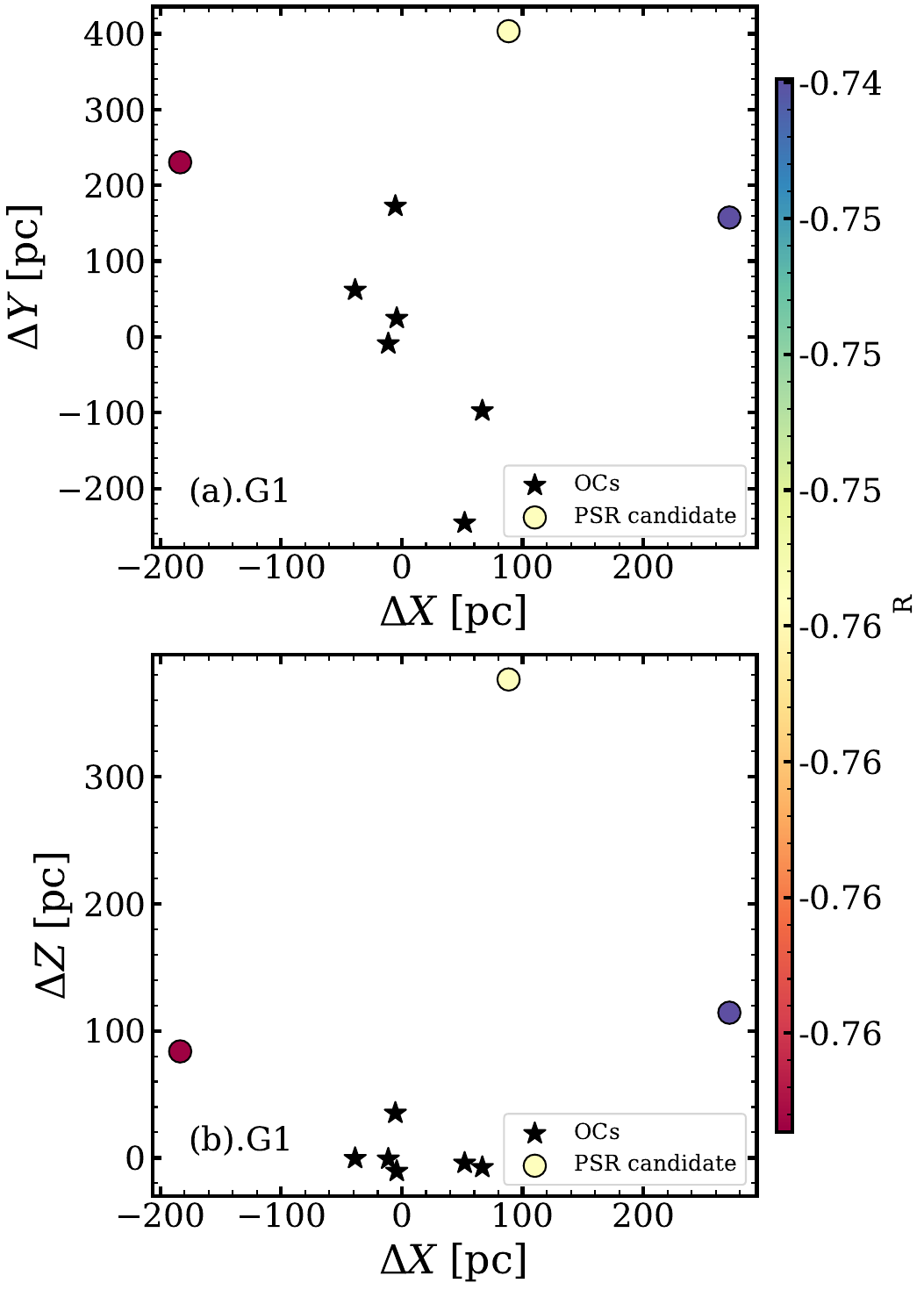}
    \includegraphics[width=0.86\linewidth]{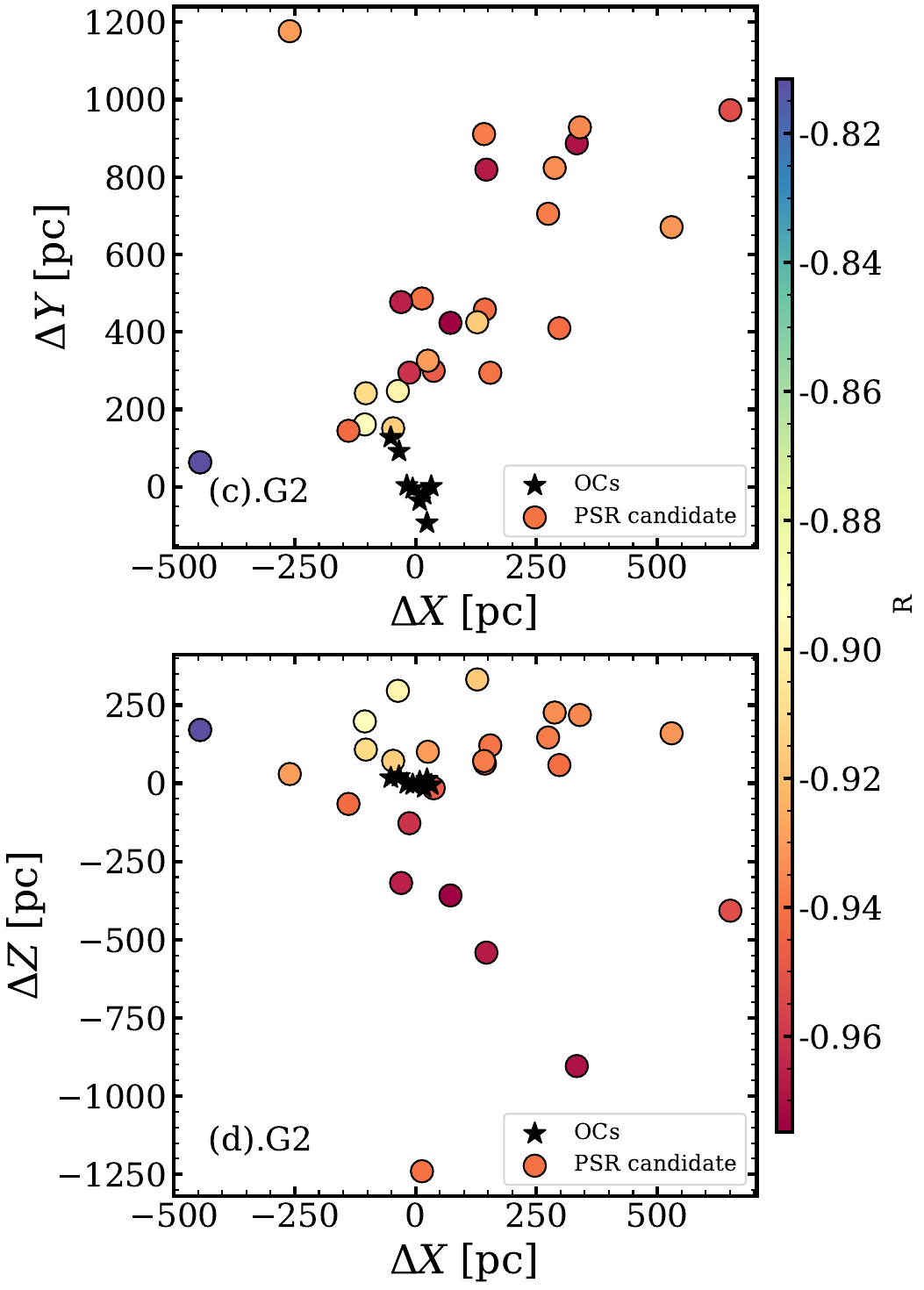}
    \caption{Relative 3D spatial distribution of PSR candidate. Panels (a)-(b): Results for G1, while panels (c)-(d) display results for G2. Black stars represent OCs, and dot symbols denote PSR candidate sources (color-coded with correlation coefficient as represented in the color bar).}
    \label{fig:psr_candidate}
\end{figure}
\section{Data and methods}\label{sec:paper1}
\subsection{Targets selection and data reduction}\label{sec:data}
We selected samples from the \textit{Gaia}-derived OC catalogs of \citealt{Cantat2020_1867} (hereafter CG20) and \citealt{Qin_2023ApJS} (hereafter Qin23), ensuring a robust and reliable dataset for our study.
CG20 provides a comprehensive catalog detailing 1,867 OCs, including member stars. 
Qin23 systematically searched for and analyzed OCs within 500 pc of the Sun using the updated \gdr{3} data. Qin23 identified a total of 324 OCs, with 101 of them newly discovered. Qin23 established a reliable and near-complete catalog of OCs in the solar neighborhood.
Our research primarily focused on a region, which centers at a Galactic longitude of 70 degrees and latitude of seven degrees, covering approximately 20 degrees in projection and distances within 700 pc. Following strict selection criteria based on position and distance, we obtained a sample of 36 OCs, including 28 OCs from Qin23 and eight OCs from CG20.

Given that the data of member star that we used originated from different versions of \gaia data (\gdr{2} and \gdr{3} ), and considering the significant improvements in \gdr{3} in terms of proper motion and parallax precision, and the substantial increase in the number of stars with radial velocity measurements from approximately seven million in \gdr{2} \citep{Katz2019} to around 34 million in \gdr{3} \citep{Katz2023}. Thus, we cross-matched all the member stars with \gdr{3} data to ensure consistent data precision and obtain more radial velocity information. We then determined the six-dimensional (6D) parameters (\RAdeg, \DEdeg, \pmra, \pmdec, \plx, RV) for the target OCs. The central positions and proper motions of the OCs were obtained using kernel density estimation. At the same time, the average line of sight velocity and distance of OCs were determined through Gaussian fitting to the RV data and the reciprocal value of the median parallax for the member stars, respectively.

To estimate the age of the OCs, we performed isochrone fitting on the color-magnitude diagram (CMD) of each target OC, using theoretical isochrones from the \gaia photometric system provided by PARSEC v1.2S \footnote{\url{http://stev.oapd.inaf.it/cgi-bin/cmd}} \citep{Bressan2012, Chen2015, Marigo2017ApJ83577M}. The best-fit isochrone was determined through visual inspection for a set of isochrones spanning a logarithmic age range of log(Age[yr]) = 6 to 10, with intervals of 0.01, and a metallicity value of solar metallicity ($Z_\odot$=0.0152, \citealt{Caffau2011SoPh268255C}). 
To obtain the mass of each star, we determined the nearest point on the best-fit isochrone and assigned the corresponding mass as the star's mass. The total OC mass was calculated by summing the masses of all member stars. We used the {\tt Astropy} package \citep{Astropy2013, Astropy2018} to calculate Galactocentric Cartesian coordinates, which were used to visualize our results in three-dimensional (3D) spatial distribution.
For these calculations, we adopted the Sun's position as (X, Y, Z) = [-8, 0, 0.015] kpc and its velocity as (U, V, W) = [10, 235, 7] \kms\ in Galactocentric coordinates \citep{Bovy2015}.

\subsection{Search method}\label{sec:method}
We searched for close correlations between nearest-neighbor pairs in 3D spatial, 3D velocity, and age within each OC group. Specifically, a target OC was considered a candidate member of a group when a nearest-neighbor OC existed within 100 pc \citep{Conrad2017, Liu_2019, Casado2021} and their age difference was less than 30 Myr. Furthermore, to increase the likelihood of a common origin, nearest-neighbor pairs were required to have a 3D velocity difference of less than 10 \kms. Additionally, we considered the velocity direction. The velocity vectors of the member OCs had to be consistent, with their movement directions confined to the same quadrant. After meeting all these criteria, both OCs in the nearest-neighbor pair were classified as potential members of the same OC group. This process was systematically repeated for each OC within the studied area.
To ensure coherence among the member OCs within an identified group, a final verification was conducted to confirm that the 3D velocity difference between any two member OCs was less than 10 \kms\ and that their velocity vectors were consistent.

Based on these strict criteria, we identified four distinct OC groups comprising 20 member OCs, which we designated as G1, G2, G3, and G4. Detailed properties of these OC groups are presented in Table~\ref{tab:group_info}, and parameters of OC are shown in both Table~\ref{tab:OC_xzy} and Table~\ref{tab:oc-para}. Among these OC groups, group G1 consists of six OCs: ASCC 105, OCSN 152, OCSN 113, OCSN 117, Roslund 5, and UPK 82. Group G2 includes eight OCs: OCSN 16, OCSN 18, Stephenson 1, UPK 101, UPK 64, UPK 72, UPK 78, and UPK 83. Groups G3 (OCSN 13, OCSN 14, and UPK 70) and G4 (OCSN 22, OCSN 23, and RSG 5) each comprise three OCs.

\section{Result and discussion}\label{sec:Discussion}

\subsection{Kinematic structures of OC groups}
The member OCs of the primordial OC group form within the same GMC, inheriting the kinematic properties of their parent cloud \citep{Elmegreen+2000}. Therefore, the member OCs within the OC group should have similar kinematic signatures. The spatial distribution and tangential velocity ($v_t$) of member OC within the four OC groups are depicted in Fig.~\ref{fig:lb_Vt}. These four OC groups each occupy a distinct spatial region, and the $v_t$ between each OC group differs. Within each OC group, the member clusters are closely adjacent and exhibit consistent velocities, indicating that they may have formed within the same GMC. 

To further analyze the 3D distribution of the identified OC groups, we applied a Bayesian method to correct the distances for individual stars \citep{Bailer2015, Carrera2019, pang_dist}. Fig.~\ref{fig:3D_xy} shows the 3D distribution of the OC groups after distance correction. The spatial extents of G1 and G2 are approximately 200 pc, while G3 and G4 extend around 60 and 80 pc, respectively. These spatial scales are consistent with the typical size of primordial OC groups, which are several hundred pc \citep{Efremov+1978,Efremov1998}. The duration of star formation within all OC groups is less than 30 Myr (as illustrated in the color bar of Fig.~\ref{fig:3D_xy} and detailed in Table~\ref{tab:group_info}). These timescales align with the expected duration of continuous star formation events \citep{Efremov1998}. These results imply that the OCs within each group were born sequentially in their respective parent molecular clouds, indicating a common origin.

To investigate the tidal interactions between OCs, we considered the relative distances between OCs and their tidal radius ($r_{\rm t}$). According to \citealt{Innanen1972}, if the distance between star clusters is less than three times the outer radius, they may have significant mutual interaction. This criterion was further examined by \citealt{deLa2009A&A500L13D}, who used three times the mean $r_{\rm t}$ as an upper limit for selecting paired OCs. Consequently, OC pairs separated by less than 3$r_{\rm t}$ are considered to have tidal interaction. The tidal radius of each star cluster was calculated following \citep{Pinfield1998MNRAS299955P}:
\begin{linenomath*}
\begin{equation}
    \centering
    r_{\rm t} = [{\frac{G M_{\rm cl}}{2 \times (A-B)^2}}]^{\frac{1}{3}} .
    \label{equ:oc_rt}
\end{equation}
\end{linenomath*}
Here $M_{\rm cl}$ is the accumulated stellar mass for an OC, G represents the gravitational constant with value G = \Gconst\ and A (15.3 $\pm$ 0.4 \kmskpc) and B (-11.9 $\pm$ 0.4 \kmskpc) are the Oort constants \citep{Bovy2017}. 
The member stars of OCs are incomplete at faint magnitudes, and $r_{\rm t}$ may be underestimated.
The grey dashed circles in Fig.~\ref{fig:3D_xy} represent the locations of the 3$r_{\rm t}$ boundaries for each OC. 
It is obvious that part of the member OCs exhibit weak interactions with others in G1 and G2, whereas the interactions between member OCs are notably stronger in G3 and G4. 

\subsection{Simulations of OC groups dynamics}\label{sec:orbit}
To deepen our understanding of the formation and evolution mechanism of the four newly identified primordial OC groups, we employed the high-performance $N$-body simulation code {\tt PeTar} \citep{Wang2020_PeTar} to trace the trajectory of OC members from their birth to their future evolutionary states. Simulating the dynamic evolution of OCs from birth presents significant challenges due to the limited availability of direct observational data on their initial conditions, such as velocity distribution, spatial configuration, and initial mass. These constraints hinder modeling from their birth time.
Therefore, we treated each member OC within each OC group as a point mass, allowing us to focus on gravitational interactions among them. 

With this approach, we integrated the trajectories of the member OCs accordingly. During the simulation, the gravitational interactions among member OCs were considered alongside the Galactic potential. The axisymmetric Galactic potential model MWPotential2014 was used, which includes disk, bulge, and halo components. {\tt MWPotential2014} is an easy-to-use model in the {\tt galpy} \citep{Bovy2015} package, designed to describe the Milky Way's gravitational potential.
The present-day 3D positions and velocities of the member OCs, as well as the total mass of the OC (as listed in Table~\ref{tab:OC_xzy}), were used as initial input parameters for the $N$-body simulation. It should be cautious that the Galactocentric coordinate used in {\tt Astropy} follows a right-handed frame, where the Sun is located at negative $X$, and $X$ increases away from the Sun. In contrast, {\tt galpy} adopts a left-handed coordinate system, meaning the $X$ coordinate is flipped relative to the right-handed frame.
For each OC group, we integrated the trajectories of member OCs back to the birth time of the oldest cluster within the OC group. As for epochs preceding the age of a given cluster, we calculated the "pre-birth" trajectory of the cluster as an approximate indicator of the parent gas cloud motion \citep{Swiggum2024Natur}. However, these "pre-birth" trajectories should be interpreted cautiously due to the inherent uncertainties in tracing the dynamic history before cluster formation.

We visually compared the trajectories for the OC members within each group in Fig.~\ref{fig:orbit_result}. The orbits of member OCs within each group exhibit nearly circular orbits, with significant overlap among their trajectories in the $X-Y$ panel. 
Along the $Z$-axis, the OCs display periodic oscillation above and below the Galactic plane with similar periods. We defined the group size $D$ (pc) as
\begin{equation*}
    D=\frac{\sum_{i=1}^{n} d_i}{n},
\end{equation*}
where $d_i$ is the distance between the $i$-th member OC and the group center, and $n$ is the total number of members in the OC group.
Fig.~\ref{fig:orbit_result} (c) displays the $D$ as a function of time. As shown in Fig.~\ref{fig:orbit_result} (c), the member OCs within each group have been moving within a confined space in the past, which is comparable to GMC sizes (100-200 pc; \citealt{Kondo2021ApJ}). However, as noted by \citealt{Efremov+1978, Vereshchagin2022}, OC groups or OB associations can exceed the size of their parent GMCs due to expansion. It becomes clear that these OCs exhibit similar orbital motions, as can be seen by combining Fig.~\ref{fig:orbit_result} (a), (b), and (c).
This consistent kinematic pattern provides evidence that the member OCs within each group move together in a coherent manner rather than a coincidental alignment.

To further understand the spatial evolution of the OC groups, we examined the properties of their sizes, represented as $D$. It is apparent in Fig.~\ref{fig:orbit_result} (c) that the $D$ value for all four OC groups was minimal in the past and has been increasing continuously, indicating that the OC groups gradually expand as they evolve from a compact state to a more dispersed state.
We also calculated the time when the OC groups reached their minimum size, which corresponds to their dynamical traceback age and marks the moment when the OC groups began expanding \citep{Miret2024NatAs}. 
G1 had a dynamical traceback age of 27 Myr and is expected to expand beyond 100 pc in approximately 11 Myr, with 100 pc being the representative scale for identifying an OC group.
G2, G3, and G4 started expanding eight, 3.5, and five Myr ago, and they are expected to reach 100 pc in 36, 38, and 44 Myr, respectively.
The shorter dispersion timescale of G1 suggests that it will expand faster than the other OC groups as evolution proceeds. 
Currently, the sizes of G2 and G4 remain close to their most compact states, while G1 and G3 are slightly larger than their most compact phases (as presented in Table~\ref{tab:group_info}). This trend indicates that G1 and G3 have been expanding at a faster rate than G2 and G4 in the past. 

We note that the time at which each group was most compact does not match the average age of the group. This phenomenon is expected in the dynamical traceback process \citep[e.g.,][]{Kounkel2020Apj, Grossschedl2021, Miret2024NatAs,Swiggum2024Natur}. Additionally, it should also be noted that the dynamical traceback age is subject to various uncertainties, including measurement errors in the 6D parameters of OCs, the presence of sub-groups, and the neglected influence of the molecular cloud gravitational field \citep[e.g.,][]{Grossschedl2021, Swiggum2024Natur}.

\begin{table} 
\caption{General properties of OC groups.} 
\label{tab:group_info} 
\centering 
\footnotesize 
\setlength{\tabcolsep}{5.0pt} 
\renewcommand{\arraystretch}{1.5} 
\begin{tabular}{ccrcccrr} 
\hline \hline 
Group  & l & b & Age & $D_{now}$ & $D_{compact}$ & $t_{compact}$ & N \\ 
  &    deg & deg & Myr & pc & pc & Myr  &  \\ 
(1) & (2) & (3) & (4) & (5) & (6) & (7) & (8) \\ \hline
G1 & 67.45 & -0.09 & ${72.00}_{-15}^{+15}$ & 94.26 & 72.41 & -27.0 & 6 \\ 
G2  & 62.94 & 13.89 & ${37.00}_{-7}^{+5}$ & 38.98 & 36.44 & -8.0 & 8\\ 
G3 & 56.50 & 3.04 & ${22.00}_{-5}^{+6}$ & 18.43 & 9.27 & -3.5  & 3 \\ 
G4 & 81.82 & 6.73 & ${41.00}_{-1}^{+12}$ & 23.96 & 21.21 & -5 & 3  \\ 
\hline 
\end{tabular} 
\tablefoot{
Columns (2) and (3): Median galactic longitude and latitude of member OCs. 
Column (4): Median age of member OCs, with lower and upper boundaries representing the differences between the median and the minimum and maximum, respectively. 
Column (5): Present-day size. 
Column 6: Most compact size. 
Column (7): Time of most compact size. 
Column (8): Number of member OCs.}
\end{table}

\subsection{Signatures of feedback in the large scale}\label{sec:origin}
The classical triggered star formation model \citep{Elmegreen1977, Elmegreen1998ASPC, Efremov1998} posits that a prior generation of massive stars significantly influences the formation of subsequent stellar populations.
The main idea is that shocks, such as those from SN explosion or other stellar feedback processes, can compress the surrounding gas and trigger a new generation of star formation. A key signature of the feedback-driven scenario is the presence of a ring or arc of gas and dust surrounding the central massive stars \citep{Verma2023ApJ}. 

In this work, we also found a similar morphology in the $l$-$b$ spatial distribution (Fig.~\ref{fig:lb_Vt}). The ring-like and arc structures observed in G1 and G2 serve as one piece of evidence for the compression of molecular clouds by feedback activities, while these features are absent in G3 and G4. As demonstrated by \citealt{Grossschedl2021} and \citealt{Cantat2019ring}, feedback from massive stars can produce such large-scale spatial morphologies. Among the various feedback mechanisms, SN explosions play a crucial role in triggering new star formation and driving large-scale gas motions exceeding 100 pc \citep{Herrington2023MNRAS}.
Thus, considering the spatial scales involved, only SN explosions can provide the necessary energy to compress and disperse the gas, shaping the observed spatial configurations. Other forms of stellar feedback alone lack the energy needed to generate structures of this scale. 

To explore the formation mechanisms of OC groups, we sought evidence of potential supernova remnants (SNRs) that could support this hypothesis. Although the OC groups are located in the Cygnus region, the Cygnus Nebula lies farther away and is situated in the background of the OC groups. Moreover, the median ages of G1 and G2 are 72 Myr and 37 Myr, making it reasonable to assume that star formation consumed the available gas, while feedback from the OC groups likely expelled remaining gas As a result, the surrounding gas of the OC groups has dissipated, making it difficult to identify direct gaseous remnants, such as SNR or HI Superbubble that could confirm the occurrence of SN explosions.
Therefore, we attempted to define a probable spatial range for SN explosions based on theoretical constraints and searched for potential pulsar (PSR) candidate sources within this range that could be the remnants of the SN explosions responsible for triggering the formation of these OC groups.
\begin{table*}[ht]
\centering 
\caption{Properties of PSR candidates.} 
\label{tab:psr_candidate} 
\footnotesize 
\setlength{\tabcolsep}{4.0pt} 
\renewcommand{\arraystretch}{1.5} 
\begin{tabular}{ccrrrrrrrrcc} 
\hline \hline 
Group & Name & l & b & pml & pmb & dist & $v_t$ & RV & age & R & p-value \\ 
~ & ~ & \ensuremath{\text{deg}} & \ensuremath{\text{deg}} & \masyr & \masyr & \ensuremath{\text{kpc}} & \kms &\kms& \ensuremath{\text{Myr}} \\ 
(1) & (2) & (3) & (4) & (5) & (6) & (7) & (8) & (9) & (10) & (11) & (12)  \\ \hline
G1 & J0645+5158 & 163.96 & 20.25 & 7.57 & -0.90 & 1.220 & 61.09 & 15.00 & 28500.00 & -0.75 & 8.30e-02 \\ 
 & J1719-1438 & 8.86 & 12.84 & -8.21 & -7.57 & 0.336 & 24.78 & -12.00 & 11400.00 & -0.74 & 8.93e-02 \\ 
 & J1809-1943 & 10.73 & -0.16 & -13.43 & 0.12 & 3.600 & 282.35 & -321.00 & 0.03 & -0.76 & 7.68e-02 \\ \hline
G2 & B0656+14 & 201.11 & 8.26 & 21.48 & 38.55 & 0.288 & 4.63 & 40.00 & 0.11 & -0.94 & 5.04e-04 \\ 
 & B1822-09 & 21.45 & 1.32 & -14.02 & 7.30 & 0.300 & 18.10 & -7.00 & 0.23 & -0.92 & 1.42e-03 \\ 
 & B1855+09 & 42.29 & 3.06 & -6.03 & -0.08 & 1.200 & 43.54 & 10.00 & 4760.00 & -0.94 & 4.54e-04 \\ 
 & B1914+09 & 44.56 & -1.02 & -5.96 & 7.03 & 1.904 & 25.53 & 14.00 & 1.70 & -0.97 & 6.38e-05 \\ 
 & B1944+17 & 55.33 & -3.50 & -7.29 & -5.37 & 0.300 & 18.10 & -19.00 & 290.00 & -0.91 & 1.67e-03 \\ 
 & B2016+28 & 68.10 & -3.98 & -6.59 & -1.31 & 0.980 & 40.73 & -17.00 & 59.70 & -0.93 & 7.45e-04 \\ 
 & J0509+0856 & 192.49 & -17.93 & 6.52 & 2.25 & 0.817 & 23.55 & 39.00 & 14600.00 & -0.90 & 2.43e-03 \\ 
 & J0613-0200 & 210.41 & -9.30 & 10.05 & -3.10 & 0.780 & 54.13 & 48.00 & 5060.00 & -0.94 & 6.49e-04 \\ 
 & J0614-3329 & 240.50 & -21.83 & 2.00 & -0.07 & 0.630 & 8.11 & 47.00 & 2840.00 & -0.92 & 1.37e-03 \\ 
 & J0645+5158 & 163.96 & 20.25 & 7.57 & -0.90 & 1.220 & 61.09 & 30.00 & 28500.00 & -0.97 & 9.40e-05 \\ 
 & J0711-6830 & 279.53 & -23.28 & -17.77 & -11.30 & 0.106 & 10.08 & 34.00 & 5840.00 & -0.94 & 5.13e-04 \\ 
 & J0737-3039A & 245.24 & -4.50 & -3.72 & -2.31 & 1.100 & 15.71 & 51.00 & 204.00 & -0.81 & 1.45e-02 \\ 
 & J0737-3039B & 245.24 & -4.50 & -3.72 & -2.31 & 1.100 & 15.71 & 51.00 & 49.20 & -0.81 & 1.45e-02 \\ 
 & J0900-3144 & 256.16 & 9.49 & -2.19 & 0.54 & 0.890 & 12.05 & 46.00 & 3600.00 & -0.96 & 1.44e-04 \\ 
 & J1045-4509 & 280.85 & 12.25 & -7.82 & 1.67 & 0.340 & 11.81 & 32.00 & 6700.00 & -0.95 & 3.37e-04 \\ 
 & J1411+2551 & 33.38 & 72.10 & -4.40 & 2.37 & 1.131 & 30.33 & -16.00 & 10400.00 & -0.97 & 3.90e-05 \\ 
 & J1536-4948 & 328.20 & 4.79 & -7.49 & 2.12 & 0.978 & 17.70 & 40.00 & 2300.00 & -0.93 & 8.09e-04 \\ 
 & J1751-2857 & 0.65 & -1.12 & -7.47 & 4.17 & 1.087 & 31.33 & 33.00 & 5530.00 & -0.94 & 5.67e-04 \\ 
 & J1801-1417 & 14.55 & 4.16 & -7.95 & 8.02 & 1.105 & 22.22 & 26.00 & 10800.00 & -0.97 & 8.44e-05 \\ 
 & J1843-1113 & 22.05 & -3.40 & -3.72 & 0.25 & 1.260 & 27.03 & 25.00 & 3060.00 & -0.89 & 2.76e-03 \\ 
 & J1853+1303 & 44.87 & 5.37 & -3.32 & 0.15 & 2.083 & 40.49 & 33.00 & 7440.00 & -0.94 & 4.51e-04 \\ 
 & J1905+0400 & 38.09 & -1.29 & -8.23 & 0.03 & 1.064 & 52.07 & 8.00 & 12200.00 & -0.94 & 5.66e-04 \\ 
 & J1911+1347 & 47.52 & 1.81 & -4.65 & 0.90 & 1.365 & 33.95 & 11.00 & 4330.00 & -0.93 & 8.13e-04 \\ 
 & J1918-0642 & 30.03 & -9.12 & -8.53 & 3.73 & 1.111 & 44.46 & 11.00 & 4720.00 & -0.93 & 6.91e-04 \\ 
 & J2032+4127 & 80.22 & 1.03 & -2.37 & 1.97 & 1.330 & 6.62 & -13.00 & 0.20 & -0.94 & 4.69e-04 \\ 
 & J2145-0750 & 47.78 & -42.08 & -12.60 & 3.73 & 0.714 & 43.51 & -40.00 & 8540.00 & -0.95 & 2.52e-04 \\ 
\hline 
\hline 
\end{tabular} 
\tablefoot{Column (2): Pulsar name. Columns (3) and (4): Galactic longitude and latitude position. Columns (5)–(6): Proper motion. Column (7): Best estimate of the pulsar distance. Column (8): Tangential velocity. Column (9): Radial velocity. Column (10): Spin-down age. Column (11): Pearson correlation coefficient between member OCs' age and their separation from the PSR candidate. Column (12): the p-value associated with the Pearson correlation coefficient.}
\end{table*}

\subsection{Modeling supernova explosion site}\label{sec:SNe}
The formation of linear cluster sequences has long been predicted as a consequence of triggered star formation \citep{Elmegreen1977, Whitworth1994, Elmegreen_11_EAS}.  Previous studies have provided observational evidence showing that stars form in spatial and temporal correlations, which supports the model of triggered star formation \citep[e.g.,][]{Zucker2022Nature, Ratzenb+2023, Verma2023ApJ, posch2024clusterchains}.
Specifically, studies of star formation influenced by SN explosions have shown that star formation does not proceed chaotically but exhibits a notable age gradient radiating outward from the explosion site. This pattern reveals older clusters positioned closer to the SN explosion sites, providing a theoretical framework to predict potential SN explosion sites around the birthplace of the two OC groups under study. 
However, due to the limited number of members in G3 and G4, establishing a robust age gradient relationship is challenging, and there is no obvious evidence of SN explosion signatures. Therefore, G3 and G4 are excluded from further discussion.

To identify potential SN explosion locations around the birth locations of OC groups, we established a relative coordinate system ($\Delta{X}, \Delta{Y}, \Delta{Z}$) centered on each OC group. Using simulation results from {\tt PeTar}, we traced the relative distribution of the member OCs in the birth time of the oldest cluster in each group.
We systematically sampled the 3D space surrounding each OC group using a step size of 10 pc. For each sampling point, we applied the least-squares fitting method to calculate the correlation coefficient between OC age and its separation from that sampling point. Finally, the correlation strength across the sampled space was constructed.
The results shown in Fig.~\ref{fig:SNe-mock} indicate a spatially convergent region where an SN explosion likely occurred. As can be seen in Fig.~\ref{fig:age-Sep}, there exists a clear relationship between the ages of the member OCs and their separation from the most probable  SN explosion site, with a distinct tendency for older OCs to be closer to the  SN explosion site and younger ones to be farther away. This spatial arrangement implies an inside-out star formation scenario for the OC groups and supports the feedback-driven model (\citealt{Efremov+1978}). The age difference between G1 and G2 implies that each OC group was formed in a separate episode, likely triggered by different SN explosions.   
As highlighted by \citealt{Stahler2004book}, individual SNR typically spans a few tens of pc. In contrast, larger regions range from \(10^2\) to \(10^3\) pc and are known as supershells. This expansive structure often results from the combined effects of multiple SNe.
Numerous observational studies support this theory. For instance, \citealt{Ehlerov+2018} found that a few tens of SN explosions create the supershell, and several OCs born in the supershell wall exhibit an age sequence along the wall. The star formation in both the Local Bubble \citep{Zucker2022Nature, Swiggum2024Natur} and Orion Complex \citep{Grossschedl2021, Kounkel2020Apj} is influenced by multiple SN explosions that may have been continued over time. 
The shockwaves generated by these SN explosions expel gas, forming shell where a new generation of star formation occurs.

Consistent with previous studies, we propose a likely star formation scenario in which several SN explosions in a short period triggered the formation of G1 and G2. In this scenario, shockwaves generated by these SNe compressed the surrounding interstellar medium into expanding shell, which subsequently fragmented and collapsed, forming the observed OC groups. It is noteworthy that the assumptions in this study are based on idealized models, and we did not account for the influence of other external factors on the gas complex, such as feedback from previously formed open clusters or the influence of galactic differential rotation. These factors might influence gas dynamics and gas dissipation processes.
However, unlike previous studies, we found that the gas density around our OC groups is lower, likely due to a significant portion of the gas consumed by star formation. Nevertheless, with the help of SN remnants, such as PSRs, we can infer where SN explosions may have occurred.

\subsection{Pulsar candidate identification}\label{sec:psr}
Pulsars are remnants of SN explosions and provide a unique opportunity to explore the history of such explosive events. 
According to the standard neutron star formation scenario \citep{Scheck2006AA457}, PSRs may have been ejected following SN explosions, a theory supported by multiple studies. For example, \citealt{Wright1979Natur} was among the first to calculate pulsar orbits,  identifying a pair of PSRs originating from a former binary system. Similarly, \citealt{Vlemmings2004ApJ610} discovered a pulsar pair with a common origin in the Cygnus superbubble based on proper motion and parallax measurements. Furthermore, \citealt{Kounkel2020Apj} found two PSRs that could be traced back to SN events in the Orion Complex. All these studies have demonstrated that the birthplaces of PSRs can be determined by tracing their trajectories \citep{Kounkel2020Apj, Vlemmings2004ApJ610, Kirsten2015, Bobylev2008AstL346}. 

Building on previous studies, we aimed to explore the correlation between PSR birthplaces and the predicted SN explosion regions to determine any potential associations with OC groups. 
For this purpose, we conducted an extensive analysis of the ATNF Pulsar Catalogue \footnote{\url{https://www.atnf.csiro.au/research/pulsar/psrcat}}, which provides detailed information on approximately 3700 known PSRs, including parameters such as position, proper motion, distance, and spin-down age, among others. Currently, reliable estimates of parameters such as distance and proper motion components are available for only a small subset of PSRs. In our study, PSRs lacking proper motion and distance data were excluded, leaving a final sample of 607 PSRs. 

Since the radial velocity (RV) of the PSR sample is unavailable, RV is treated as a free parameter when determining the trajectory of PSR. To estimate the RV of PSRs, \citealt{Bobylev2008AstL346} utilized various RV (e.g., +200 or −200 \kms), while \citealt{Vlemmings2004ApJ610} considered three different RV (−300, 0, and +300 \kms). However, both studies sampled too few RV values, limiting the accuracy of the results. In our study, we aim to implement a more precise and comprehensive RV sampling to improve the accuracy of PSR birthplace determination. Assuming that the tangential velocity ($v_t$) is equivalent to the RV, we could determine the RV range of the sample.
We calculated the $v_t$ by combining the proper motion and the distance. We analyzed the $v_t$ distribution of the pulsar sample and restricted the sample to within 1$\sigma$. As a result, we found that nearly 89\% of the PSRs have $v_t$ values below 400 \kms. Leveraging this distribution, we performed an exhaustive sampling for each PSR, covering a range of $\pm 400$ \kms\ with an increment of 1 \kms, assigning 800 potential RV values for each PSR.
To accurately trace back to the birthplace of a PSR, it is essential to account for its age. It is also important to note that differences between chronological ages and spin-down ages can be significant \citep{Bobylev2008AstL346}, making the exact determination of PSR ages challenging. Therefore, we assumed the age of each PSR to be equal to that of the oldest cluster in the associated OC group. Using the {\tt galpy} software package, we traced the orbits of PSRs to determine their birthplaces. Specifically, we determined the only RV for each PSR at which it most closely encountered these OC groups following the method of \citealt{Bobylev2008AstL346}.

For PSRs located within the predicted SN explosion region, we further examined the correlation between the ages of member OCs and the separation of PSRs from the OCs. We discarded PSRs that show weak correlations. To further refine our selection, we constrained the PSR candidate scale height to $Z$ < 1 kpc, consistent with the scale height of the Galactic thick disk, which ranges from approximately 500 to 1900 pc \citep{Vieira2023Galax1177V}. Additionally, we limited the difference between RV and $v_t$ to within 50 \kms. 
As a result of these refined selection parameters, we obtained three PSRs whose birthplaces coincide with the predicted SN explosion regions around the birth location of group G1. In the vicinity of G2, 26 PSRs were found with birthplaces falling within the predicted SN explosion regions (see Fig.~\ref{fig:psr_candidate}). Therefore, we imply that these PSRs could be remnants of the SN explosions. The specific properties of these candidate PSRs are detailed in Table~\ref{tab:psr_candidate}.

The presence of PSRs associated with OC groups provides additional observational evidence for SN explosions in this region, further supporting the scenario of supernova-triggered star formation for G1 and G2. However, due to the limited availability of reliable information, we consider this a possible hypothesis.
It should be noted that our estimates of PSR ages are approximate. When accounting for the spin-down age, no PSRs near G1 are comparable to the oldest member cluster. G2 has one PSR whose age is equivalent to its oldest member OC. If the remnant is a black hole, it is even more difficult to detect. The spatial scales are comparable for G1 and G2, however, the number of PSR candidates in their vicinity differs significantly. This discrepancy is partly due to the incompleteness of the PSR sample and partly attributed to the potential influence of black holes, as previously mentioned.

\section{Summary}\label{sec:summary}
In this study, we comprehensively identified and analyzed OC groups utilizing high-precision astrometric data from the \gaia mission.  
We report four new primordial OC groups: G1 (six members), G2 (eight members), G3 (three members), and G4 (three members), by examining each OC group candidate across 3D spatial position, 3D velocity, and age. The main results are summarised as follows:
\begin{enumerate}
      \item Each OC group occupies a distinct region in phase space. The spatial proximity and kinematic coherence of member clusters within each group strongly suggest a common origin within the same giant molecular cloud. The age distribution and spatial extent of the OC group are consistent with the timescale of continuous star formation events and the typical scale of the primordial OC group, indicating the sequential formation of member clusters.
      \item We employed the $N$-body simulation code {\tt PeTar} to trace the evolutionary paths of the open clusters from their formation to their future states. The results imply a gradual dispersal of member OCs in each group, ultimately leading to their evolution into independent entities. 
      \item The star formation history of groups G1 and G2 supports a likely hypothesis: independent SN explosion events triggered the formation of these two groups. We further determined potential SN explosion sites through a negative correlation between the ages of OCs and their separation from the SNe source. These patterns are reminiscent of the classic feedback-driven star scenario formation \citep{Efremov+1978}, suggesting an important role for feedback on the formation of the OCs.
      \item  We yield three candidate PSRs potentially associated with the formation of group G1 and 26 candidates for group G2 by tracing the orbital trajectories of PSRs, based on the ATNF Pulsar Catalogue. We propose that these PSRs could be remnants of the SN explosions, which strengthens the inference that SN explosions triggered the sequential star formation in OC groups. However, due to the limited availability of reliable information, we consider this to be a possible hypothesis.
   \end{enumerate}
Our results provide support for supernova-triggered star formation in OC groups, offering valuable insights into the interactions between stellar evolution and cluster formation within the Milky Way. Additionally, this supports the hierarchical star formation model, further highlighting the complex multi-scale processes involved in OC formation.

\begin{acknowledgements}
We are grateful to an anonymous referee for valuable comments, which have improved the paper significantly.
This work is supported by the National Natural Science Foundation of China (NSFC) through grants 12090040, 12090042, 12073060, 12333008, and 12288102. 

Y.Z. acknowledges the support from the science research grants from the Chinese Academy of Sciences (CAS) "Light of West China" Program (No. 2022-XBQNXZ-013), the Natural Science Foundation of Xinjiang Uygur Autonomous Region (No. 2022D01E86), Central Guidance for Local Science and Technology Development Fund (No. ZYYD2025QY27), the National Key R\&D Inter governmental Cooperation Program of China (No. 2023YFE0102300) and the Tianshan Talent Training Program with No. 2023TSYCCX0101.  

J.Z. would like to acknowledge the National Key R\&D Program of China No.2019YFA0405501， the Youth Innovation Promotion Association CAS, the science research grants from the China Manned Space Project with NO. CMS-CSST-2021-A08, the Science and Technology Commission of Shanghai Municipality (Grant No. 22dz1202400), the Shanghai Science and Technology Program (Grant No.20590760800), and Sponsored by Program of Shanghai Academic/Technology Research Leader.

X.M. acknowledges support from the International Centre of Supernovae, Yunnan Key Laboratory (No. 202302AN360001), and Yunnan Fundamental Research Projects (Nos. 202401BC070007 and 202201BC070003).

K.W. acknowledges support from the DFG grant Sp 345/24-1.

G.L. thanks Ruqiu Lin for the helpful discussions and Qiuyi Luo for assisting with checking the gas data.
This work has made use of data from the European Space Agency (ESA) mission \gaia, processed by the Gaia Data Proces-Processingalysis Consortium (DPAC,\url{https://www.cosmos. esa.int/web/gaia/dpac/consortium}).
In addition to those cited in the main body of the text, this work made use of the open source Python packages Astropy \citep{Astropy2013, Astropy2018}, SciPy \citep{scipy}, matplotlib \citep{matplotlib}, NumPy \citep{numpy}, scikit-learn \citep{skilearn}, galpy \citep{Bovy2015} and PeTar \citep{Wang2020_PeTar}.

\end{acknowledgements}

\bibliographystyle{aa}
\bibliography{reference}
\begin{appendix}
\section{Parameter of OC members}
\begin{table}[!ht]
\centering
\caption{Basic parameters of the OCs in this study.}
\label{tab:oc-para}
\footnotesize 
\setlength{\tabcolsep}{6.0pt}
\renewcommand{\arraystretch}{1.5}
\begin{tabular}{rrrrrrrrrrrr}
\hline
Group & Cluster & l & b & pml & pmb & dist & Rv & logt & DM & E(B-V)&  $r_{\rm t}$ \\ 
     ~ & ~ & \ensuremath{\text{(deg)}} & \ensuremath{\text{(deg)}} & (\masyr) & (\masyr) & \ensuremath{\text{(pc)}} & (\kms) & ~ & \ensuremath{\text{(mag)}} & \ensuremath{\text{(mag)}} & \ensuremath{\text{(pc)}} \\ 
     (1) & (2) & (3) & (4) & (5) & (6) & (7) & (8)& (9) & (10) & (11) & (12) \\ \hline \hline
        G1    & ASCC 105     & 62.88 & 2.10  & -0.68 & -2.03 & 558.45 & -16.75 & 7.85 & 9.30 & 0.18   & 7.05  \\ 
        ~ & OCSN 152     & 66.19 & 5.86  & 0.03  & -2.86 & 435.56 & -20.13 & 7.91 & 8.63 & 0.11   & 6.31     \\
      ~& OCSN 113     & 68.10 & -0.36 & 0.61  & -3.86 & 372.73 & -18.53 & 7.82 & 8.28 & 0.09   & 9.14        \\
      ~& OCSN 117     & 71.28 & -0.05 & 2.43  & -4.37 & 329.65 & -18.20 & 7.76 & 7.83 & 0.05   & 7.54        \\
      ~& Roslund 5    & 71.39 & 0.14  & 0.12  & -2.39 & 545.16 & -18.01 & 7.87 & 9.04 & 0.14   & 7.82        \\
      ~& UPK 82       & 63.26 & -0.54 & -0.69 & -3.04 & 540.86 & -16.26 & 7.93 & 8.94 & 0.11   & 6.25        \\
        \hline
       G2    & OCSN 16      & 62.47 & 13.52 & -2.76 & -2.47 & 372.45 & -16.50 & 7.54 & 8.04 & 0.07   & 7.12  \\
      ~& OCSN 18      & 63.92 & 15.40 & -2.54 & -1.69 & 371.13 & -17.58 & 7.56 & 8.06 & 0.07   & 5.67        \\
      ~& Stephenson 1 & 66.89 & 15.34 & -2.31 & -2.21 & 356.57 & -18.42 & 7.58 & 8.01 & 0.08   & 7.06        \\
      ~& UPK 101      & 70.21 & 15.64 & -0.80 & -1.80 & 351.17 & -18.99 & 7.59 & 7.89 & 0.06   & 8.27        \\
      ~& UPK 64       & 56.47 & 10.33 & -4.65 & -1.31 & 430.20 & -15.96 & 7.60 & 8.95 & 0.26   & 7.19        \\
      ~& UPK 72       & 57.98 & 10.21 & -4.35 & -1.45 & 456.91 & -16.62 & 7.63 & 8.83 & 0.18   & 6.09        \\
      ~& UPK 78       & 61.02 & 16.65 & -3.39 & -1.54 & 372.82 & -16.79 & 7.56 & 8.06 & 0.06   & 5.45        \\
      ~& UPK 83       & 64.40 & 12.67 & -2.03 & -3.23 & 352.72 & -16.00 & 7.48 & 7.99 & 0.06   & 4.99        \\
        \hline
        G3    & OCSN 13      & 55.66 & 2.05  & -5.06 & -4.54 & 499.51 & -9.71  & 7.35 & 8.55 & 0.23   & 9.21 \\
      ~& OCSN 14      & 56.93 & 3.20  & -4.39 & -3.92 & 514.11 & -4.23  & 7.25 & 8.47 & 0.14   & 5.21        \\
      ~& UPK 70       & 56.28 & 6.24  & -4.90 & -2.52 & 493.50 & -12.06 & 7.45 & 8.52 & 0.20   & 6.41        \\
        \hline
        G4    & OCSN 22      & 80.94 & 13.19 & 3.43  & -2.09 & 306.14 & -13.69 & 7.72 & 7.35 & 0.03   & 6.22  \\
      ~& OCSN 23      & 84.88 & 6.73  & 3.31  & -1.39 & 345.67 & -6.87  & 7.61 & 7.91 & 0.08   & 7.45        \\
      ~& RSG 5        & 81.75 & 6.05  & 3.34  & -2.21 & 337.20 & -7.12  & 7.60 & 7.89 & 0.07   & 7.86        \\  \hline
    \end{tabular}
    \begin{minipage}{0.92\textwidth}
    \textbf{ Notes.} Column (1): OC group name in this work. Column (2): Cluster name in this work. Columns (3) and (4): Median galactic longitude and latitude of member stars. Columns (5) and (6): Median proper motion of member stars. Column (7): Reciprocal value of the median parallax for the member stars. Column (8): Radial velocity determined by Gaussian fitting. Column (9): Cluster age determined by the isochrone fit. Column (10): Distance modulus determined by the isochrone fit. Column (11): Reddening determined by the isochrone fit. Column (12): Tidal radius of cluster.
    \end{minipage}
\end{table}
\end{appendix}
\end{CJK*}
\end{document}